      [1999/12/01 v1.4c Il Nuovo Cimento]
\documentclass{cimento}

\usepackage{graphicx}  
\usepackage{amsmath,amssymb,amsopn,bm}
\newcommand{\beq}{\begin{equation}}
 \newcommand{\eeq}{\end{equation}}
 \newcommand{\beqa}{\begin{eqnarray}}
 \newcommand{\eeqa}{\end{eqnarray}}
 \DeclareMathOperator{\im}{Im}
 
\title{Hadronic and electromagnetic probes of hot and dense matter in a 
Boltzmann+Hydrodynamics model of relativistic nuclear collisions}

\author{E.~Santini\from{ins:x}, B.~B\"{a}uchle\from{ins:x}\from{ins:y}, H.~Petersen\from{ins:z}, J.~Steinheimer\from{ins:x}, M.~Nahrgang\from{ins:x} \atque
M.~Bleicher\from{ins:x}\from{ins:y}}
\instlist{\inst{ins:x} Institut f\"ur Theoretische Physik, Goethe-Universit\"at,
 Max-von-Laue-Str.~1, D-60438 Frankfurt am Main, Germany
         \inst{ins:y} Frankfurt Institute for Advanced Studies (FIAS), Ruth-Moufang-Str.~1,D-60438 Frankfurt am Main, Germany
         \inst{ins:z}Department of Physics, Duke University, Durham, NC 27708, United States}
\PACSes{24.10.Lx, 25.75.-q, 25.75.Dw, 25.75.Cj}
\begin{document}

\maketitle

\begin{abstract}
We present recent results on bulk observables and electromagnetic probes 
obtained using a hybrid approach based on the 
Ultrarelativistic Quantum Molecular Dynamics
 transport model with an intermediate hydrodynamic stage
for the description of heavy-ion collisions at AGS, SPS and RHIC energies.
After briefly reviewing the main results for particle multiplicities, elliptic flow, 
transverse momentum and 
rapidity spectra, we focus on photon and dilepton emission from hot and dense 
hadronic matter.
\end{abstract}

\section{Introduction}
The investigation of nuclear matter under extreme conditions is one of the 
major research topic of nuclear and high energy physics. 
Experimental information about the properties of hot and dense strongly 
interacting systems is sought by analysing high energy collisions of heavy 
nuclei. To link specific experimental observables to the 
different manifestations and, eventually, phases of the strongly interacting 
matter, a detailed understanding of the dynamics of the heavy-ion reactions 
is essential. 

Numerous observables, such as hadronic and electromagnetic probes, 
their dynamical pattern and some specific correlations they exhibit, 
have been and are currently investigated in detail experimentally. 
All these observables are generally connected in a non-trivial way. 
A challenging task, in this respect, is a meaningful modelling of 
the heavy ion collision. 
Microscopic (transport) and macroscopic (hydrodynamical) models attempt 
to describe the full time evolution of the heavy-ion reactions
and have played, in their various realizations, an important role in 
the interpretation of the experimental results over the last decades.

Recently, a third class of models, so-called ``hybrid approaches'', 
has been developed. 
Hybrid approaches combine the advantages of transport approaches that are well suited to deal with the non-equilibrium initial and final states, with those of an intermediate hydrodynamic evolution, where, \emph{e.g.}, the equation of state (EoS) is an explicit input and phase transitions can be easily implemented. 
Such approaches were proposed 10 years ago
\cite{Dumitru:1999sf,Bass:1999tu} and have since then been employed for a wide
range of investigations
\cite{Teaney:2001av,Socolowski:2004hw,Nonaka:2005aj,Hirano:2005wx}. The hybrid
approach discussed here is based on the integration of a hydrodynamic evolution
into the Ultra-relativistic Quantum
Molecular Dynamics (UrQMD) transport approach \cite{Petersen:2008kb}. 
This integrated hybrid approach, called UrQMD v3.3 \footnote{Website of the UrQMD Collaboration http://urqmd.org.}, has been 
applied by the Frankfurt group to the investigation of various 
hadronic observables in the broad energy range from $E_{lab}=2-160$ GeV \cite{Petersen:2008dd,Li:2008qm,Petersen:2009vx,Petersen:2009mz,Petersen:2009zi,Petersen:1900zz} and, 
recently, of electromagnetic probes (photons and dileptons) \cite{Bauchle:2009ep,Santini:2009nd,Baeuchle:2010ym,Baeuchle:2010yq,Santini:2010in}, which have the 
unique feature of being sensible to the \emph{whole}  time evolution of the 
system. Note that in the hybrid model emission of virtual and real photons 
from the QGP phase can be 
explicitly accounted for provided that an EoS with a 
QGP phase is used for the hydrodynamical evolution.
This constitutes a great advantage with respect to the pure hadronic transport 
models, where such emission cannot be easily implemented.
In this proceeding, we briefly survey some of the results obtained 
both for the hadronic and electromagnetic sector.

\section{A hybrid approach to heavy-ion collisions}
Hybrid models generally schedule three dynamical stages: during the initial stage a microscopic transport scheme carries the incidentally colliding nuclei towards a stage that determines the initial conditions for the relativistic hydrodynamic equations of motion. At this stage one assumes local equilibration and the fireball is subsequently followed by a hydrodynamic evolution until the description is handed over to a final-state kinetic transport description. This late stage automatically furnishes a continuous freeze-out process, an important improvement compared to the otherwise employed prescription of an instantaneous freeze-out.
Below, we want to briefly summarize the specific realization 
of the hybrid approach by the Frankfurt group and 
refer the reader to \cite{Petersen:2008kb} for further details.

During the first stage of the evolution the particles are described as a 
purely hadronic cascade within UrQMD. The coupling to the hydrodynamical 
evolution proceeds when the two Lorentz-contracted nuclei have 
passed through each other.
At this time, the spectators continue to propagate 
in the cascade and all other hadrons
are mapped to the hydrodynamic grid.  
Subsequently, a $(3+1)$ ideal hydrodynamic evolution is performed using
the SHASTA algorithm \cite{Rischke:1995ir,Rischke:1995mt}. 
The hydrodynamic
evolution is later merged into the hadronic cascade. 
 Two possible transition criteria
procedures where tested. The first is the isochronous freeze-out (IF). In this
approach, all hydrodynamic cells are mapped onto particles at the same time,
once the energy drops below five times the ground state energy density in all
cells. The second criterium is called gradual freeze-out (GF). In this approach
transverse slices, of thickness 0.2 fm, are transformed to particles 
whenever in all
cells of each individual slice the energy density  drops below
five times the ground state energy density. 
The employment of such gradual transition allows to 
obtain a rapidity independent transition temperature 
without artificial time dilatation effects \cite{Steinheimer:2009nn}.
When merging, the hydrodynamic fields are transformed to particle
degrees of freedom via the Cooper-Frye equation. 
The created particles proceed in their evolution in
the hadronic cascade where final state interactions and decays of the particles occur within the UrQMD framework.

\section{Hadronic probes: a brief review}

\begin{figure}[h]
 \center
\includegraphics[width=.82\textwidth]{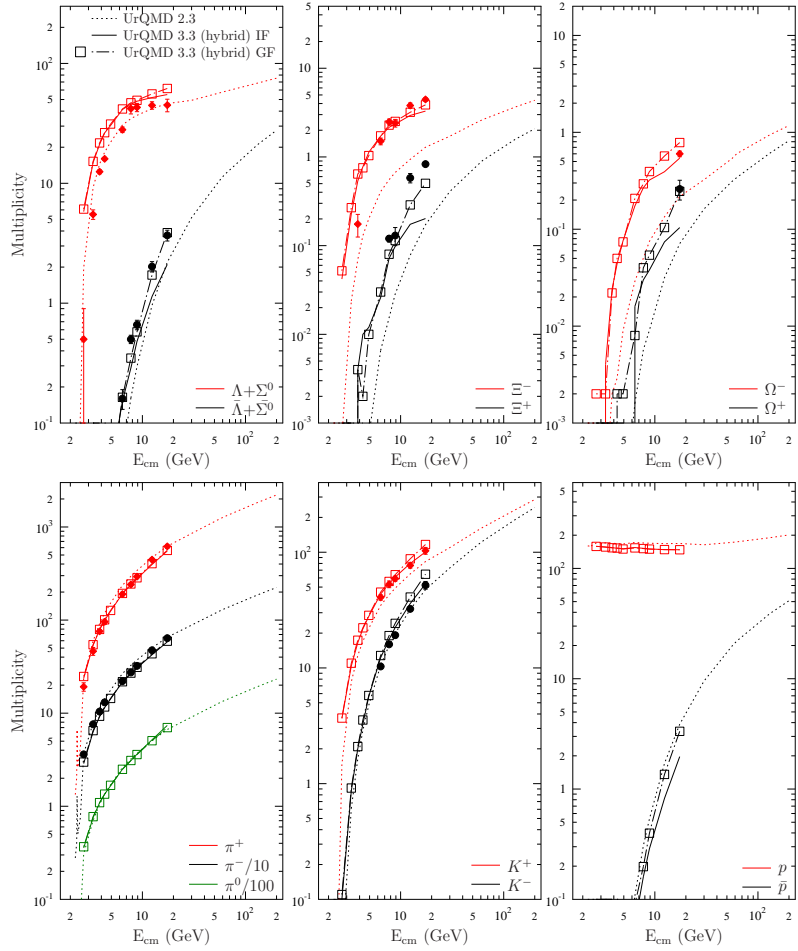}
\caption{(Color Online) Excitation functions of the $\Lambda$, $\Xi$, $\Omega$, $\pi$, $K$
and $p$ multiplicities ($4\pi$) in central ($b<3.4$ fm) Au+Au/Pb+Pb collisions.
The results for the hybrid model with isochronous freeze-out (full lines), the
hybrid model with gradual freeze-out (squares) and pure UrQMD-2.3 (dotted
lines) are compared to experimental data (full symbols) 
from various experiments \cite{Klay:2003zf,Pinkenburg:2001fj,Chung:2003zr,:2007fe,Afanasiev:2002mx,
Anticic:2003ux,Mitrovski:2006js,Alt:2008qm,Blume:2004ci,Afanasiev:2002he,Alt:2004kq}.
\label{fig:exc}} \end{figure}

Compared to treatments solely throughout by kinetic transport, 
as, \emph{e.g.}, by UrQMD, such hybrid strategies provide a well pronounced 
sensitivity of the transverse flow on the hydrodynamic part of the 
evolution \cite{Li:2008qm,Petersen:2009mz,Santini:2009nd}. 
Among others, they enhance the production of strange 
particles \cite{Petersen:2008dd,Petersen:2009zi}. 
In Fig. \ref{fig:exc}, the excitation functions of the $\Lambda$, $\Xi$,
$\Omega$, $\pi$, $K$ and $p$ yields \cite{Graf:2010pm} are shown . The enhancement of the
multiplicities for all particles with strange content compared to the non-hybrid
results (dotted lines) can be clearly observed.

Investigations of the longitudinal dynamics via rapidity distributions of 
various hadron species have shown that the latters are not too 
sensitive to the details of the dynamics for the hot and dense stage 
\cite{Petersen:2008dd}.  
The rapidity distributions for, \emph{e.g.}, $\pi^-$  and $K^+$ 
at three  different energies ($E_{\rm lab} = 11,40$ and 
$160A~$GeV) have been analysed in Ref.~\cite{Petersen:2008dd}, where  it was 
shown that the general shape of the distribution resulting in the 
hybrid approach 
is very similar to the one obtained by pure cascade calculations  and 
in line with the experimental data \cite{Akiba:1996xf,Afanasiev:2002mx}.  

An additional outcome of such models is that 
elliptic flow is found to increase towards collider energies 
\cite{Petersen:1900zz} compared to pure cascade calculations. 

\section{Electromagnetic probes}

Electromagnetic probes, such as photons and lepton pairs, 
are penetrating probes of the hot and dense matter. 
Once created these particles pass the collision zone essentially 
without further interaction and can therefore mediate valuable information 
on the electromagnetic response of the strongly interacting medium.

In the hybrid approach, emission of real and virtual (dileptons) photons is treated as follows. 
During the locally equilibrated 
hydrodynamic stage the production of lepton pairs and direct photons is 
described by radiation rates for a strongly interacting medium in thermal 
equilibrium (so-called ``thermal'' photons and dileptons).  
In  the evolution stage that precedes or follows the 
hydrodynamical phase, the emission is performed as typically done in 
cascade approaches. More specifically, dileptons emission is calculated 
employing the time integration method  
that has long been 
applied in the transport description of dilepton emission 
(see \emph{e.g.} \cite{Schmidt:2008hm}), whereas direct photon emission is calculated 
according to cross-sections for direct photon production from various channels. 
We will briefly discuss the main aspects of photons and dilepton emission in the hybrid approach here below. 
For further detail, the reader is referred to \cite{Bauchle:2009ep,Santini:2009nd,Santini:2010in}.

\subsection{Photons}

The most important hadronic channels for the production of direct photons
are $\pi\pi\rightarrow\gamma\rho$ and $\pi\rho\rightarrow\gamma\pi$~\cite{Kapusta:1991qp}, which
both are implemented in the transport, as well as in the hydrodynamic phase.
The cross-sections for cascade-calculations are taken from Kapusta {\it et
al.}~\cite{Kapusta:1991qp}, while the rates used for the hydrodynamic
description have been parametrized by Turbide {\it et
al.}~\cite{Turbide:2003si}.  Since no thermal partonic interactions are
modelled in UrQMD, emission from a QGP-medium is only taken into account in
the hydrodynamic part of the model.  Several minor hadronic channels are
only implemented in one of the two models, such as strange channels (\emph{e.g.} 
$K\pi\rightarrow\gamma{}K^\ast$) which are only present in the hydrodynamic
calculations, and $\eta$-channels (\emph{e.g.}\ $\pi\eta\rightarrow\gamma\pi$)
which are only present in the transport calculations. Earlier investigations
with this model have shown those channels to provide only minor contributions 
to the overall spectrum of direct photons. 
In the Quark-Gluon-Plasma, the rate used is taken from 
Ref.~\cite{Arnold:2001ms}, where convenient parametrizations for the 
contribution of
$2\leftrightarrow2$, bremsstrahlung- and annihilation-processes are given. The complete
list of channels and a detailed explanation of the calculation procedure is 
provided
in~\cite{Bauchle:2009ep}.
\begin{figure}\center
\includegraphics[width=0.6\textwidth]{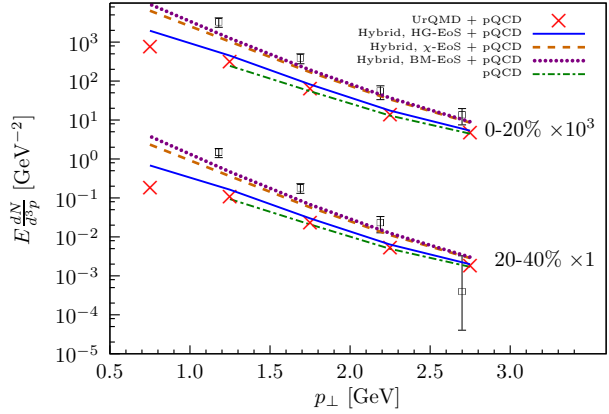}
 \caption{(Color Online) 
Comparison of direct photon spectra (open squares)  for central (0-20~\%)
and mid-central (20-40~\%) Au-Au collisions at  $\sqrt{s_{NN}}=200$ GeV~\cite{:2008fqa} to 
different calculations: a) cascade (red crosses), b) hybrid model with
HG-EoS (blue solid line), c) $\chi$-EoS (orange dashed line), and d) BM-EoS
(violet dotted line).  The
 contribution from initial
 pQCD-scatterings~\cite{Gordon:1993qc,:2008fqa} has been added to
 all spectra. The spectra from central collisions have been scaled by a
 factor of $10^3$ to enhance readability.
 }
 \label{fig:auau:200}
\end{figure}

In Fig.~\ref{fig:auau:200} a comparison between direct photon spectra 
from hybrid model calculations and data from the PHENIX collaboration~\cite{:2008fqa} for central (0-20\%) 
and mid-central (20-40\%) Au+Au collisions 
at $\sqrt{s_{NN}}=200$ GeV is shown. 
The experimental data were obtained by extrapolating the dilepton yield to 
zero invariant mass~\cite{:2008fqa}. The hybrid model calculations were 
performed using a hadron gas EoS describing a non-interacting gas of free 
hadrons \cite{Zschiesche:2002zr} (HG-EoS, blue solid lines), a chiral EoS 
 that follows from 
coupling a chiral hadronic SU(3) 
Lagrangian with a PNJL-type quark-gluon description \cite{Steinheimer:2010}
($\chi$-EoS, orange dashed lines), 
and bag model EoS that exhibits a strong first order phase 
transition between a
Walecka type hadron gas and massless quarks and gluons \cite{Rischke:1995mt} 
(BM-EoS, violet dotted lines). Pure cascade calculations are indicated by red crosses. 
All calculated spectra include the $\langle N_{\sf coll} \rangle$-scaled 
prompt photon contribution taken from ~\cite{Gordon:1993qc,:2008fqa}.
We observe that in both centrality-bins, the direct photon spectra obtained 
with the BM-EoS and $\chi$-EoS, which include a phase transition to a 
deconfined state of matter, are significantly higher than the hadronic 
HG-EoS-calculations. Similar enhancement of photon production due to QGP 
emission was found already at SPS energies~\cite{Bauchle:2009ep}.

\subsection{Dileptons}
Invoking vector meson dominance the emission rate of lepton pairs from a 
strongly interacting medium can be related, at low invariant masses, 
to the spectral properties of the 
vector mesons, 
with the $\rho$ meson giving the dominant contribution. 
The thermal dilepton rate reads then \cite{Rapp:1999ej}:
\beq
\frac{d^8 N_{ll}}{d^4 x d^4 q}=-\frac{\alpha^2m_\rho^4}{\pi^3 g_\rho^2}\frac{L(M^2)}{M^2}f_B(q_0;T)
\im D_\rho(M,q;T,\mu_B) \, ,  
\label{rate}
\eeq
where $\alpha$ denotes the fine structure constant, 
$M^2=q_0^2-q^2$ the dilepton invariant mass squared, $f_B$ the Bose 
distribution function 
(for a moving fluid this must be substituted with the J\"{u}ttner function), 
and 
$L(M^2)$ a lepton phase space factor that quickly approaches one above the 
lepton pair threshold. 
The electromagnetic response of the strongly interacting medium is then 
encasted in $\im D_\rho(M,q;T,\mu_B)$, 
the imaginary part of the in-medium $\rho$ meson propagator,
\beq
D_\rho(M,q;T,\mu_B)=\frac{1}{M^2-m_\rho^2-\Sigma_\rho(M,q;T,\mu_B)}.
\label{rhoprop}
\eeq
Over the years, strong evidence has been accumulated pointing to the 
conclusion that the inclusion of the in-medium contributions to the $\rho$ 
meson self energy is mandatory for a proper description of the low invariant 
mass region of the dilepton spectra 
(see e.g. Ref.~\cite{Hendrik_wish}), typically 
characterized by the emerging of an enhancement with respect to the standard 
hadronic cocktail~\cite{Michele_wish}.
A nice aspect of the hybrid approach is that the hydrodynamic stage 
allows for a transparent inclusion of the in-medium spectral function 
of the vector meson, conceptually problematic in transport calculations, 
in analogy to fireball model calculations \cite{vanHees:2007th,Ruppert:2007cr}. 
An advantage with respect to the latters is the use of a dynamical model 
for the description  of the heavy-ion collisions, which might help to shed 
some light on dynamical aspects as dilepton radial flow or similar. 

In this application, the self-energy
contributions taken into account are $\Sigma_\rho=\Sigma^0+\Sigma^{\rho\pi}+\Sigma^{\rho N}$,
where $\Sigma^0$ is the vacuum self-energy and $\Sigma^{\rho\pi}$ and $\Sigma^{\rho N}$
denote the contribution to the self-energy due to the direct
interactions of the $\rho$ with, respectively, pions and nucleons of the
surrounding heat bath. The self-energies have been 
calculated according to Ref. \cite{Eletsky:2001bb}, where they were 
evaluated in terms of empirical 
scattering amplitudes from resonance dominance at low energies and 
Regge-type behaviour at high energy.

\begin{figure}
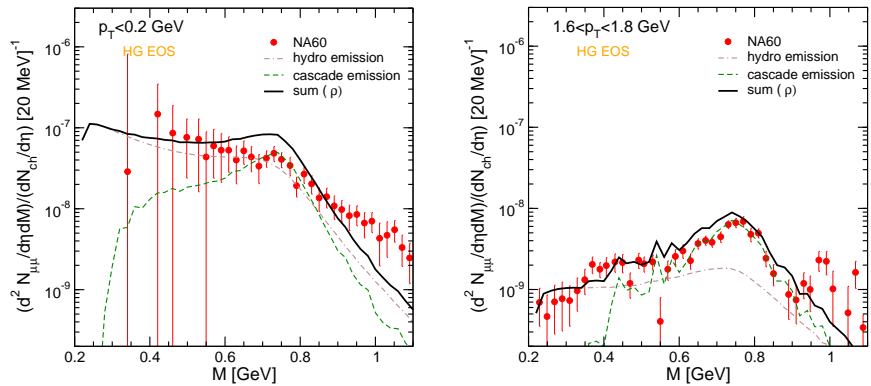

  \begin{center}
  \includegraphics[width=0.4\textwidth]{proc.HGim.am.pt.le.0.2.eps}
\hspace{.4cm}
  \includegraphics[width=0.4\textwidth]{proc.HGim.am.1.6.pt.1.8.eps}
  \end{center}
  \caption{\label{na_60_HGim.am.pt.le.0.2} (Color Online) Left panel: Acceptance-corrected invariant 
 mass spectra of the excess dimuons in In-In collisions at 158$A$ 
 GeV for transverse pair momenta $p_T<0.2$ GeV, 
 compared to hybrid model calculations based on thermal radiation from 
 in-medium modified $\rho$ meson spectral function (dotted-dashed line) and non-thermal cascade emission (dashed line). 
The sum of the two contributions is depicted by the full line. Experimental data from Ref.~\cite{Arnaldi:2008fw}. Right panel: Same as in the left panel, but for the transverse momenta window  $1.6<p_T<1.8$ GeV.}
\end{figure}

In Fig.~\ref{na_60_HGim.am.pt.le.0.2} 
hybrid model calculations are compared to recent acceptance-corrected 
NA60 data 
\cite{Arnaldi:2008fw}. 
The calculations have been exemplary performed using a hadron gas equation 
of state for the hydrodynamical evolution.
This is enough to  point out qualitatively the 
main features of the present approach. 
More systematic studies and discussions will be 
presented elsewhere \cite{Elviranext}.
We observe that the cascade emission dominates the invariant mass region 
around the vector meson peak 
for both low and intermediate transverse pair momenta $p_T$. 
At low $p_T$ 
(left panel of Fig.~\ref{na_60_HGim.am.pt.le.0.2}), 
the  very low invariant masses, $M<0.5$ GeV, 
are filled by the thermal 
radiation with in-medium spectral function. 
The sum of both contributions, however, 
leads to an overestimation of the vector meson 
peak region at low transverse momenta of the dilepton pair. 
The reason for the discrepancy might partially lie on the specific 
spectral function used here 
and/or, presumably more severely, on the eventual presence of not yet negligible residual 
in-medium modification 
of the $\rho$ meson spectral function during the cascade stage, 
that are here neglected. 
With increasing  $p_T$ (right panel of Fig.~\ref{na_60_HGim.am.pt.le.0.2}), 
dilepton emission at very low invariant masses 
is reduced and the total spectra are almost completely determined by 
the cascade emission.  
Indeed, thermal emission had already shown 
discrepancies for $p_T>1$  GeV \cite{vanHees:2007th}, pointing to the 
necessity to account for non-thermal contributions. In the present approach, 
the latters appear quite naturally.

Finally, we would like to mention that, 
in analogy to the direct photons calculations,  
the present approach easily allows 
for the inclusion of dilepton emission from QGP, which is thought to play 
a role in the intermediate mass region of the dilepton spectra. 

\appendix

\acknowledgments
E.S. thanks the organizers for the invitation and for providing partial local support. 
This work was supported by the Hessen Initiative 
for Excellence (LOEWE) through the Helmholtz International Center for FAIR 
(HIC for FAIR) and in part by U.S. department of
Energy grant DE-FG02-05ER41367.
B.\ B. gratefully acknowledges support from the
Deutsche Telekom Stiftung, the Helmholtz Research School on Quark Matter
Studies and the Helmholtz Graduate School for Hadron and Ion Research.
H.P. acknowledges a Feodor Lynen fellowship of
the Alexander von
Humboldt foundation.
We thank the Center for Scientific Computing for providing 
computational resources.

\bibliography{biblio}

\end{document}